\documentclass{article}

\usepackage[dvips]{epsfig}
\usepackage{rotating}

\begin{document}
 
\title{The Measure of Compositional Heterogeneity in DNA Sequences
Is Related to Measures of Complexity
\vspace{0.1in}
\author{Wentian Li\\
{\small \sl Laboratory of Statistical Genetics, 
Rockefeller University, Box 192}\\
{\small \sl 1230 York Avenue, New York, NY 10021}\\ 
} 
\date{to be published in {\sl Complexity} (1997)}
} 
\maketitle    
\markboth{\sl Li, heterogeneity}{\sl Li, heterogeneity}



DNA sequences store the complete genetic information of a biological
organism. Understanding the ``genetic language" in DNA sequences
is the ultimate goal of the Human Genome Project, which will have a
profound impact on biology, medicine, and human society \cite{code}.
In one sense, the genetic language written in DNA sequences is
simpler than the English language because it is composed 
of only four letters A, C, G, and T, representing the four 
nucleotides (also referred to as bases).
In another sense,  DNA sequences are more complex than English 
because of their extreme long length, which allows more combinations 
of letters, thus more ``words". So how complex can a given DNA sequence be? 

Besides the question of genome complexity, whose measure ranges 
from the total length of the DNA sequences to the total number of 
genes produced in a genome, there is also a question of how
complicated the {\sl symbolic text} of a DNA sequence is.
This sequence complexity question requires a measurement of the statistical 
features in the arrangement
of the four nucleotides. We now know that the base composition and 
correlations among neighboring bases are not the major contributing factors 
to the sequence-wide pattern. It is the spatial
heterogeneity of the base composition \cite{bernardi-rev1,bernardi-rev2}
or the long-range correlations \cite{wli-rev1,wli-rev2} that largely
shapes the complexity of the whole sequence.

At first, it is thought that the heterogeneity in DNA sequences is simple.
In the bacteriophage lambda sequence, for example, the spatial difference
of base composition is ``black-and-white": the C+G density is higher on 
the left half of the sequence, and lower on the right half. Some people
think that the bacteriophage lambda sequence is a good representative
of all DNA sequences and that the long-range correlation can be completely 
explained by this simple heterogeneity \cite{karlin}.

As one of the people who first observed the long-range correlation 
in DNA sequences \cite{wli92,wli-kaneko},  it was clear to the author that
this proposition was not correct.  In fact, the surprise in our first
observation of the long-range correlation in DNA sequences was not 
the long-range {\sl per se}, but a special type of 
long-range correlation called ``1/f spectra" or ``1/f noise". 
Simple heterogeneity could lead to a deviation from the random sequences 
(``white noise" or ``white spectra"), but does not automatically lead to 
``1/f spectra".  Extra features besides simple heterogeneity are needed to 
explain this special type of long-range correlation.

Some recent developments in the study of compositional heterogeneity
of DNA sequences enable us to address this issue in a satisfactory
fashion \cite{pepe96,pepe97}. With respect to the base composition, 
a DNA sequence can be homogeneous, heterogeneous in a simple way, or 
heterogeneous in a complex way. The complex heterogeneity is 
characterized by the ``domains within a domain" 
phenomenon \cite{wli-rev1,wli-rev2,pepe96,pepe97}. 
For these sequences, whether a region is homogeneous or not is 
only relative. Interestingly, the measure of base composition
heterogeneity coming out of the study of long-range 
correlation in DNA sequences are intrinsically related to measures 
of complexity in the study of complex systems. This connection is the topic
to be addressed in this Commentary.

\subsection*{Partitioning a heterogeneous DNA sequences into homogeneous
domains}

First things first: we need to partition a heterogeneous DNA sequence into 
two relatively homogeneous subsequences. In the field of information
theory,  a quantity called Jensen-Shannon distance \cite{lin} can be used
to measure the distance between two statistical distributions. This
Jensen-Shannon distance is defined as the difference between 
the entropy (another well-known quantity in information theory, as
well as in statistical physics) calculated from the whole system,
and the weighted sum of entropies calculated from the subsystems.
The Jensen-Shannon distance is successfully applied to the DNA
sequences for the purpose of partitioning the sequence \cite{pepe96,pepe97}. 

One first calculates the Jensen-Shannon distance $D$ for each possible partition 
point $i$ along the DNA sequence.  This $D(i)$ function is plotted in Fig.1 
for the DNA sequence from the first chromosome of budding yeast 
(whose academic name is {\sl Saccharomyces cerevisiae}). This sequence 
contains 230,208 nucleotides. The higher the value of the Jensen-Shannon
distance $D(i)$ at a given point $i$, the bigger the difference of
the two subsequences as partitioned at point $i$, and the more ideal to choose that point 
to partition the sequence. In Fig.1, the highest point of $D(i)$ is 
actually reached at one telomere region - the end of the chromosome, 
and the second highest point at another telomere region. What it tells 
us is that both telomeric regions are quite different from the rest of 
the sequence,  with respect to base composition.

Besides the two telomeric regions, the third highest point of
$D(i)$ in Fig.1 is near $i \approx 189,000$. The fourth highest
point is near $i \approx 27,000$, etc. Overall, there are several 
other places where we can partition the sequence and the resulting
base composition difference between the two subsequences is large.

The Jensen-Shannon distance for a random sequence scrambled from the
yeast sequence is also calculated (Fig.1). Although there are 
also ups-and-downs, the average value of $D(i)$ is at least 
10 times lower than that for the yeast sequence. These ups-and-downs in 
$D(i)$ for the random sequence are purely random fluctuations.

What about the bacteriophage lambda sequence? Its $D(i)$ function
is shown in Fig.2 (again, a random sequence is included for comparison). 
The sequence length is 48,502 nucleotides. There is an
unambiguous optimal partition point around $i \approx 22,000$ 
which maximizes the Jensen-Shannon distance between the two subsequences. 
Just like the boundary separating a black and a white region, 
moving away from the boundary gradually mixes some black with the 
dominantly white region or white with the dominantly black region,
and the Jensen-Shannon distance $D(i)$ monotonically decreases. This is
exactly what happens in Fig.2, indicating an easily describable 
heterogeneity structure in the bacteriophage lambda sequence.

\subsection*{Domains within domains}

When the partitioning process \cite{pepe96} 
is recursively applied to each subsequence
of an already-partitioned sequence, sequences with simple heterogeneity
are expected to behave differently from sequences with complex heterogeneity. 
If the sequence is a simple ``black and white" type - 
``black" on one side, ``white" on the other, further partitioning  is not
expected to reveal new structures in the subsequences. On the other 
hand, if there are sub-domains within a domain, sub-sub-domains within 
a sub-domain, etc., the recursive partitioning can go on much longer
and further down to the smaller length scales.

Since even homogeneous random sequences can have small differences between
any two regions, we might want to distinguish the partitioning due to true 
heterogeneity and that due to a random fluctuation.
A significance level $s$ can be set as the cut off point. For example, 
if $s$ is set at 99.9\%, partitioning is halted when the Jensen-Shannon 
distance is not as large as would be expected by a 0.1\% chance due 
to pure random fluctuation. When recursive partioning is finally halted,
all delineated subsequences are true homogeneous regions with a probability 
of 99.9\%.

At this point, a final Jensen-Shannon distance can be calculated
which adds up the distributional differences between each
subsequence-pair in each stage of the partitioning (with a certain
weight so the partitioning of a shorter subsequences contributes
less to the final distance than longer subsequences) \cite{pepe97}. 
The result thus obtained is called ``compositional complexity" $D^*(s)$
in \cite{pepe97}. It is a function of the significance level 
$s$, and a $*$ is used to indicate that each partition 
point at each stage is optimally chosen to maximize the 
Jensen-Shannon distance at that stage of the partition \cite{pepe97}.
Also note that $D^*(s)$ is a measure of the distance among many
distributions, whereas $D(i)$ is a distance between two distributions.

This final Jensen-Shannon distance as a function of the significance level
for these four sequences, two DNA sequences and two scrambled random sequences,
is plotted in Fig.3. We
can see that at the same $s$ value, $D^*(s)$ for the yeast sequence 
is always larger than that of the bacteriophage lambda sequence. 
This again supports our early conclusion that there is a higher 
degree of heterogeneity in the yeast sequence than in the bacteriophage lambda
sequence. As for the random sequences, these behave as homogeneous sequences
when $s$ is large - with a very small number of domains.  When $s$ is reduced, 
the random fluctuation leads to spurious heterogeneity. The random 
sequence corresponding to the bacteriophage lambda sequence even overtakes 
the original sequence as $s$ is reduced, meaning that the bacteriophage lambda 
sequence is very similar to a random sequence once the simple heterogeneity 
is removed - a point debated so heatedly in the literature and so easily 
illustrated by this $D^*(s)$ plot!

\subsection*{The perspective of spectral analysis}

The difference between DNA sequences being homogeneous, heterogeneous 
in a simple way, and heterogeneous in a complex way, can be elegantly shown
by the $D^*(s)$ plot. Here I want to comment that these differences 
can also be shown (though not so elegantly) by the traditional 
spectral analysis.

Power spectrum is a technique used to represent the correlation structure
in a sequence according to wavelength (or, frequency $f$ which 
is the inverse of the wavelength). The power at a given
frequency, $P(f)$, is the contribution from that frequency
component to the total variance of the fluctuation in the sequence.

A random sequence lacks correlation at any length scale,
and the contribution to the total variance of fluctuation
in the sequence from each frequency component is the same. When
the $P(f)$ of a random sequence is plotted,  it is flat.
Using an analogy to visible light, since the color white takes
equal contribution from colors of all frequencies, a random
sequence with a flat power spectrum is also known as
``white noise" (see, e.g., \cite{gardner}).

Now what about sequences with simple or complex heterogeneity?
The answer is not obvious. Let me calculate the $P(f)$'s for our
four sequences and show these in Fig.4.
The two random sequences have flat power spectra as expected.
Both yeast chromosome 1 and the bacteriophage lambda sequence
deviate from the flat spectrum. But can we distinguish
simple and complex heterogeneity by the power spectra?

We actually can. More discussion can be found in \cite{wli-rev1,wli-rev2}. 
The proposition is that DNA sequences with simple heterogeneity exhibit 
$1/f^2$ power spectra, whereas those with complex heterogeneity frequently
exhibit $1/f$ power spectra (for a readable account of $1/f$ and $1/f^2$ 
power spectra, see \cite{gardner,schroeder}).  Power spectra of other
shapes are of course possible, but not common in DNA sequences. Also,
the grouping of all possible spectra into only $1/f$ and a $1/f^2$ is
a simplification: considering the case of a $1/f^{1.5}$ power spectrum,
for example. The reason that such simplified picture is presented is
to emphasize the importance of the difference between $1/f$ and a $1/f^2$
power spectra.

In Fig.4,  a $1/f^2$ and a $1/f$ function (these are
straight lines in the double-logarithmic plot) are shown as reference 
functions which can be compared with the power spectra from the
two DNA sequences.  Indeed, the bacteriophage lambda sequence and the
yeast chromosome exhibit different spectra: $1/f^2$ spectrum for
the former and $1/f$ spectrum for the latter.

\subsection*{Measure of heterogeneity complexity}

Our original goal was to distinguish DNA sequences with simple and 
complex heterogeneity, and the introduction of $D^*(s)$ seems to 
be able to accomplish this task. Does $D^*(s)$ have anything
to do with measures of complexity in the field of complex systems
studies?

We first need to clarify what it is meant by a measure of complexity.
In the most general framework, a measure of complexity of a task is 
{\sl any measure that characterizes the difficulty in accomplishing 
that task} (e.g. \cite{wli91}). A measure of complexity of an object is 
{\sl  a measure of complexity of a task performed on that object}. 
Describing an object using a specific language with a specific
set of vocabularies, for example, is a task performed on that object. 
All the following 
examples can be considered a measure of complexity of a symbolic 
sequence: the length of the shortest description of a sequence
(algorithmic complexity \cite{komo}); the length of the shortest
description of the regularities in a sequence 
(effective complexity \cite{gellman}); the time required 
to reproduce a sequence from a short, if not the shortest,
description, or the time consumed in finding this short
description (logical depth \cite{depth}), etc.

When describing an object, one can describe every detail
(a strong description) or only the non-random regularities of
the object (a weak description). Correspondingly, there can be
strong and weak versions of a measure of complexity.  In describing 
the heterogeneity of base composition in a DNA sequence, we clearly
describe a specific regularity, thus the weak version. If we use
the length of a description of the heterogeneity in a DNA sequence
as the measure of complexity, the question is: is $D^*(s)$ such
a measure?

$D^*(s)$ mainly contains two pieces of information: the total number 
of homogeneous domains and the magnitude of base composition
differences among these domains. Increasing either one of these,
$D^*(s)$ is also increased. With some exceptions, the length
of a description of the heterogeneity in a sequence increases
with the total number of domains. One exception is the
case of perfectly periodic domain structures, which nevertheless
is rarely applicable to DNA sequences.

The magnitude of the differences among domains does not necessarily
contribute to the length of a description of the heterogeneity.
However, a stronger difference between domains makes the domain
structure more convincing. We might consider a larger difference
among domains a better assurance that the number of domains obtained
is correct. So the magnitude of the differences indirectly 
contribute to the length of a description of the heterogeneity.

The most interesting common feature between $D^*(s)$ and the measure
of complexity is that they both increase with the level of details
in the description. Intuitively, details not visible to the naked
eye could be revealed by a magnifying glass. Similarly, a presumably
homogeneous domain with a higher significance level can be partitioned
to more domains when the significance level is reduced. Also,
a description working at a crude level does not describe the details
at a finer level. All these arguments point out that $D^*(s)$ and
measures of complexity are monotonic functions with the level of description: 
the smaller the $s$, the higher the $D^*(s)$; 
the finer the detail, the larger the measure of complexity.

In general, the measure of complexity must take into account 
at what level one wants to describe the object \cite{gellman}.  Random 
sequences require a long description if all details are to be described, 
but a very short one if only a rough picture is required. A measure of 
complexity for a random sequence is thus unstable with respect 
to the level of detail in the description. To my knowledge, 
very few, if any,  proposed measures of complexity actually address the
issue of the level of detail in the description. The $D^*(s)$, however,
is explicitly a function of the level $s$. Perhaps we can learn a lesson
or two from the measure of heterogeneity, $D^*(s)$, and introduce 
level-dependence explicitly to the measure of complexity.

\subsection*{Acknowledgements}

I would like to thank J\'{o}se L. Oliver for providing me with
the data used in Fig.3, J\'{o}se L. Oliver and Ram\'{o}n Rom\'{a}n-Rold\'{a}n 
for comments, and Katherine Montague for proofreading the paper.
This work is supported by a grant from National Human Genome Research 
Institute of National Institute of Health (K01HG00024).

\newpage

\newpage

\begin{figure}
\begin{center}
\epsfig{file=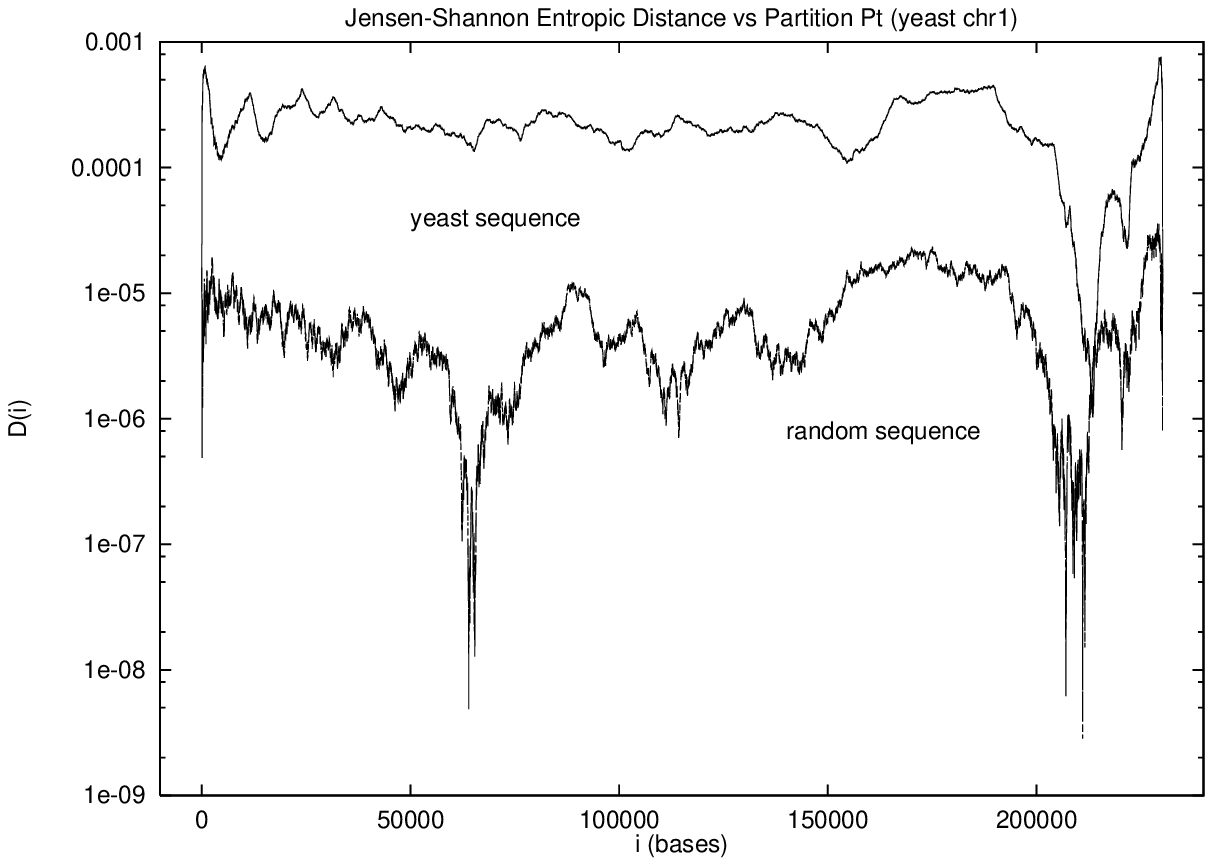}
\end{center}
\caption{
Jensen-Shannon distance $D(i)$ between the subsequence on 
the two sides as a function of the partition point $i$ 
for the DNA sequence in budding yeast chromosome 1 and 
a scrambled random sequence (same length and same base composition). 
A logarithmic scale is used for the $D$ values.
}
\label{fig1}
\end{figure}

\begin{figure}
\begin{center}
\epsfig{file=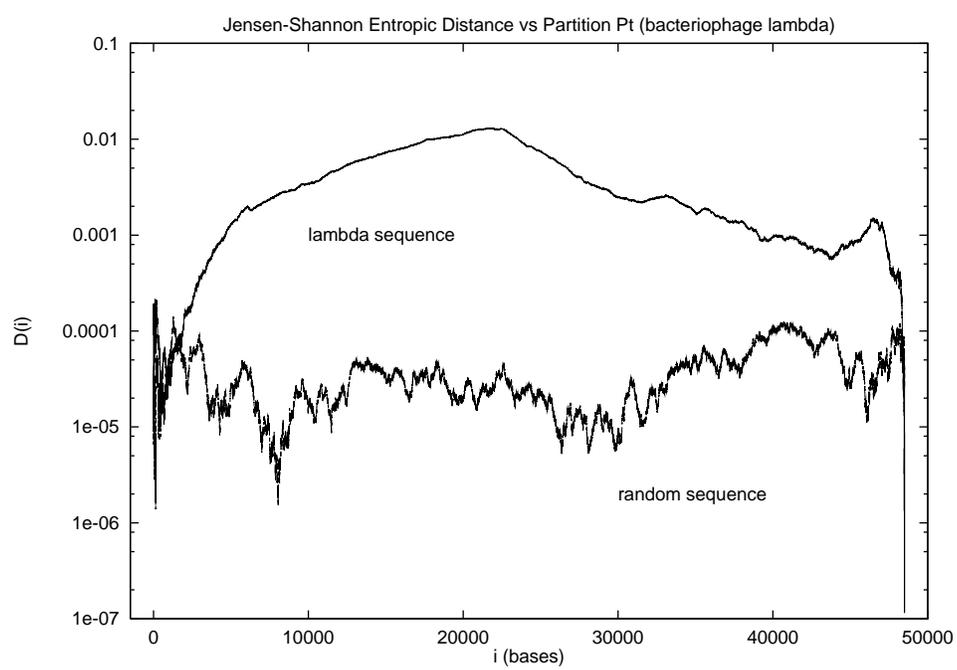}
\end{center}
\caption{
Similar to Fig.1 but for the bacteriophage lambda sequence.
}
\label{fig2}
\end{figure}

\begin{figure}
\begin{center}
\epsfig{file=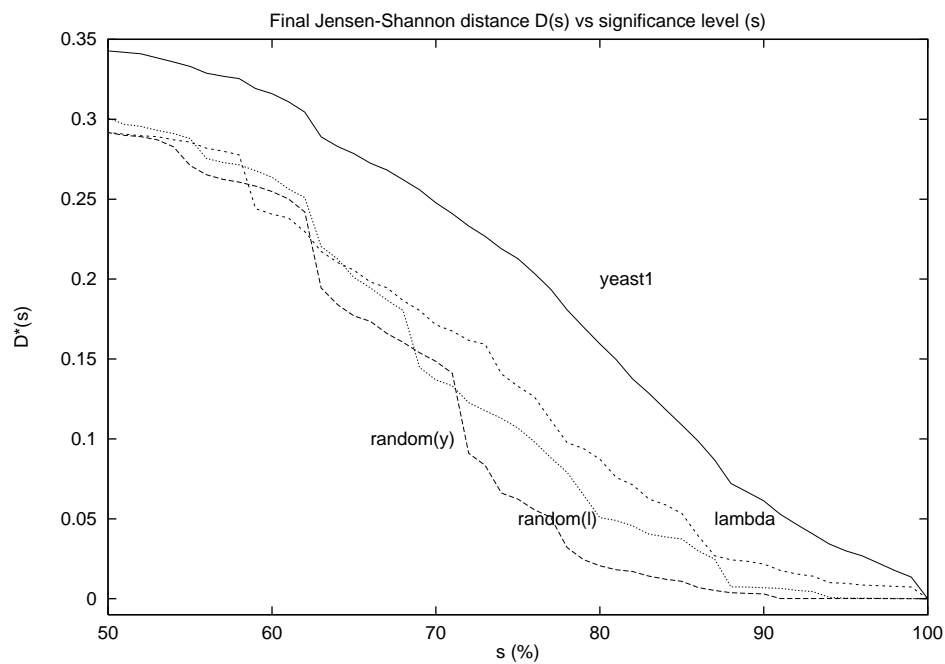}
\end{center}
\caption{
Final Jensen-Shannon distance $D^*(s)$ (``compositional
complexity")  as a function of the significance
level $s$ for yeast chromosome 1 sequence, bacteriophage lambda
sequence and the two corresponding random sequences. The larger the
value of $s$, the more difficult to partition a sequence, and
the more likely that the resulting partition reflects true
heterogeneity rather than a random fluctuation.
}
\label{fig3}
\end{figure}

\begin{figure}
\begin{center}
\epsfig{file=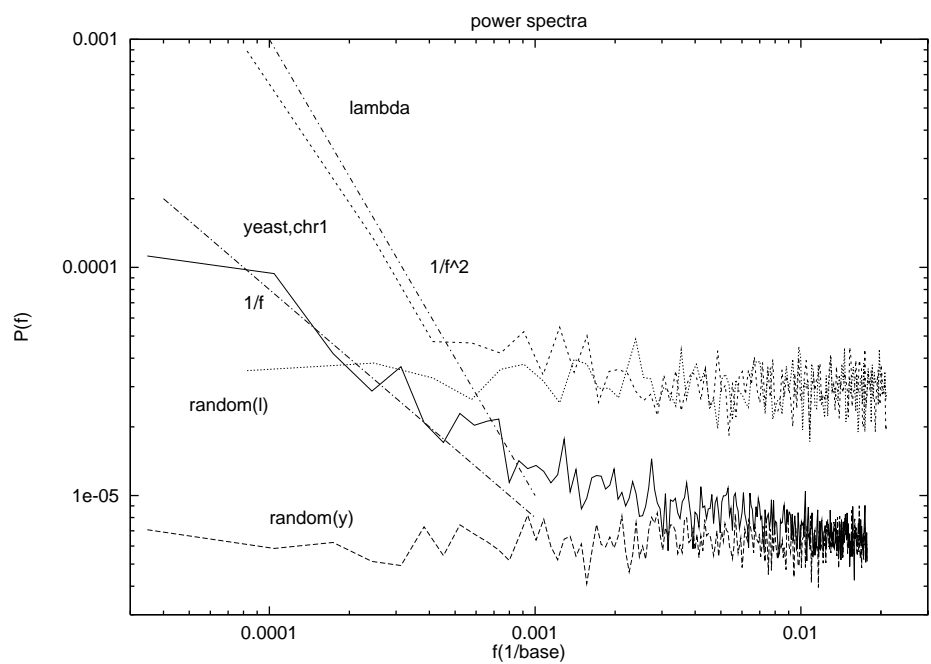}
\end{center}
\caption{
Power spectra $P(f)$ of the DNA sequence from the yeast chromosome 1,
the bacteriophage lambda sequence, and two corresponding random
sequences.  The power $P$ is plotted as a function of the frequency $f$
(both in the logarithmic scale). These power spectra are smoothed.
Two reference  lines are also shown: one represents a $1/f$ spectrum,
and the other a $1/f^2$ spectrum.
}
\label{fig4}
\end{figure}

\end{document}